\documentclass{ws-ijmpd}
\usepackage[super,compress]{cite}
\usepackage{bm}
\usepackage{hyperref}
\newcommand{\der}[2]{\frac{\partial#1}{\partial#2}}
\newcommand{\eu}{\ensuremath{\mathrm{e}}}
\DeclareMathOperator\erf{erf}
\begin{document}

\markboth{Marco Merafina and Matteo Teodori}{}

%
\catchline{}{}{}{}{}
%
\title{GENERALIZATION OF THE FOKKER-PLANCK EQUATION FOR STELLAR ORBIT DIFFUSION IN MULTI-MASS STAR SYSTEMS}

\author{MARCO MERAFINA}
\address{Department of Physics, University of Rome La Sapienza, Piazzale Aldo Moro 2, I-000185 Rome, Italy\\ marco.merafina@roma1.infn.it}
\author{MATTEO TEODORI}
\address{Istituto di Astrofisica e Planetologia Spaziali - Istituto Nazionale di Astrofisica, Via del Fosso del Cavaliere 100, 00133 Rome, Italy\\
Department of Physics, University of Rome La Sapienza, Piazzale Aldo Moro 2, I-000185 Rome, Italy\\matteo.teodori@inaf.it}


\maketitle


\begin{abstract}
We improve the standard theory of collisional stellar systems by considering the presence of a continuous mass distribution. The calculus of the diffusion coefficients is generalized and a new expression of the Fokker-Planck equation is found for multi-mass systems. A King-like distribution function, which validates the basic assumptions of most multi-mass models for Globular Clusters existing in literature, is obtained.
\end{abstract}

\keywords{Collisional systems, gravitational encounters, stellar dynamics, mass distribution}

\ccode{PACS numbers: 98.20.Gm, 98.10.+z, 98.35.Ce}


\section{\label{sec:intro}Introduction}
Globular Clusters (GCs) are collisional stellar systems, where the motion of individual stars is caused by the mean field of the other stars (referred to as the field stars) and perturbed by stellar encounters. 
The evolution of GCs is characterized by the relaxation time, the time scale over which stellar encounters become important in the dynamics of the system. Collisional systems have a relaxation time less than their age and therefore the importance of stellar encounters has a great impact on the dynamical evolution of these clusters. 
GCs also exhibit star evaporation, \cite{Ambartsumyan1938, Spitzer1940, King1958II,King62, BinneyTremaine,PiattiCarballo-Bello2020} a phenomenon which allows stars to escape the system due to the presence of tidal forces induced by the Galaxy, which impart finite escape velocities and lead to a finite boundary \cite{vonHoerner1957,King62}.  
The great importance of stellar encounters and the existence of a finite escape velocity rejects a description based on the Gaussian velocity distribution function (DF), and requires a different distribution produced by the effects of such encounters (as highlighted by many authors, see Refs.~\refcite{Chandra1960,SpitzerHarm,Michie1963}). This distribution drops to zero at a finite limiting velocity.  

On the other hand, there exist in the literature many articles which model the relaxation process by taking into account the so-called ``resonant relaxation", which applies principally in multi-star systems in the presence of a central black hole \cite{BahcallWolf1976,BahcallWolf1977,RauchTremaine1996,Alexander2017}. Nevertheless, our treatment adopts a non-resonant relaxation approach to GCs since this appears to be the main effect in the evolution of such systems. In fact already over many decades several papers on stellar dynamics in GCs have considered King-like DFs, where only the single-mass DF arises from the Fokker-Planck equation in the formulation introduced by Chandrasekhar \cite{Chandra1943}, given that the conditions for the validity of this approximation are fully satisfied. The conditions for which the multi-mass King DF often used in the literature are valid as a solution in the framework of the Fokker-Planck approximation are discussed there.  
In this view, the behavior of collisional systems is generally described by the collisional Boltzmann equation, as pointed out by Chandrasekhar in 1943 \cite{Chandra1943}
\begin{equation}
    \label{eq:CollBoltz}
   \frac{df}{dt}=\der{f}{t}+\bm{v}\cdot \bm{\nabla}f-\bm{\nabla}\varphi\der{f}{\bm{v}}=\Gamma(f)\>,
\end{equation}
where $f=f(\bm{r},\bm{v})$ is the star DF, $\varphi$ is the gravitational potential, namely the mean field produced by all the stars, while $\Gamma(f)$ is the collisional term describing the perturbations of the potential $\varphi$ due to stellar encounters.

\section{\label{sec:Single-mass solution}The single mass solution}

In self-gravitating systems like GCs, binary encounters between stars make the evaluation of the collisional term challenging. However, using the Fokker-Planck equation, which assumes a local approximation and low energy exchanges, it is possible to express the collisional term in the following form
\begin{eqnarray}
    \label{eq:Gammaf}
\Gamma(f)=-\der{}{v_i}[f(\bm{r},\bm{v})\langle \Delta v_i\rangle]
+\frac{1}{2}\frac{\partial^2}{\partial v_j\partial v_i}[f(\bm{r},\bm{v})\langle \Delta v_i\Delta v_j\rangle]\>,
\end{eqnarray}
as shown by Chandrasekhar, in 1943 \cite{Chandra1943a}. 
The convention that repeated indices are summed is observed here.
The terms $\langle \Delta v_i\rangle$ and $\langle\Delta v_i \Delta v_j\rangle$ are called diffusion coefficients and describe the time variation of the velocity and dispersion velocity due to stellar encounters.
The diffusion coefficients can be calculated from accumulation of binary stellar encounters in a homogeneous central region between a test star with mass $m$ and the field stars with mass $m_a$ \cite{Chandra1943}.
The work by Chandrasekhar was also used for studying ionized gas, in particular by Cohen, Spitzer and Routly \cite{CohenSpitzerRoutly1950}. 

Following Binney and Tremaine (see Refs.~\refcite{BinneyTremaine,Chandra1943a,Rosenbluth}), the diffusion coefficients must be inserted into the expression for $\Gamma(f)$ to re-express the collisional term to obtain the Fokker-Planck equation in the Spitzer-H\"arm form \cite{SpitzerHarm}, given by their Eq.~(1).
They seek a solution of this equation by assuming that $f(x,t)=\exp{\big({-\lambda\>{t/t_R}}\big)}g(x)$, where $\lambda$ is the evaporation rate, $t_R$ the central relaxation time (See Eq.~(3) of Ref.~\refcite{SpitzerHarm}) and $x={v}/{\sqrt{2}\sigma}$ the dimensionless velocity, with $\sigma$ the one-dimensional velocity dispersion for the field stars near the center of the cluster. 
The quantity $\lambda$ is the evaporation rate at which the stars leave the system, as a consequence of the existence of a finite escape velocity induced by tidal forces due to the gravitational potential of the Galaxy. Writing the velocity distribution as $g(x)=A\Bar{g}(x)$, where $A$ is a normalization constant, and following King, we expand $\Bar{g}(x)$ in a power series in $\lambda$, since the evaporation rate is typically very small. The approximate solution $\Bar{g}(x)$ and the evaporation rate are given by \cite{King65}
\begin{subequations}
    \label{eq:Kingm-ma}
\begin{eqnarray}
    \Bar{g}(x)&=& \frac{e^{-x^2{(m/m_a)}}-e^{-x_e^2{(m/m_a)}}}{1-e^{-x_e^2{(m/m_a)}}}\>,\label{eq:Kingm-ma_a}\\
   \lambda&=& \frac{8}{\sqrt{\pi}}\bigg(\frac{m}{m_a}\bigg)^{\frac{5}{2}}\frac{1}{e^{x_e^2{(m/m_a)}}-1} \> ,
   \end{eqnarray}
\end{subequations}
\noindent
where $x_e={v_e/\sqrt{2}\sigma}$ is the dimensionless escape velocity.
In the framework of single-mass models, it is assumed that $m=m_a$. 

Requiring that in the limit of infinite escape velocity, one recover the Maxwell-Boltzmann DF $g(x)$ with the correct normalization such that the integration over velocity gives the (central) number density $n_{0,m}$ for the test stars, the normalization constant $A$ must satisfy
\begin{eqnarray}
    A=n_{0,m}/(2\pi\sigma^2)^{3/2}\>.
    \label{eq:A_monomass}
\end{eqnarray}
Note that King in 1965 did not require this normalization and in 1966 he introduced a normalization factor $k$ (Eq.~(3) of Ref.~\refcite{King66}) which also includes the denominator in Eq.~(\ref{eq:Kingm-ma_a}). 
In order to extend the validity of the DF to the entire equilibrium configuration, we start from the form valid in the central region of the cluster and apply Jeans theorem \cite{Jeans1915} to obtain \cite{King66}
\begin{equation}
    g(r,v)=
    ke^{-{[\varphi(r)-\varphi_0]/\sigma^2}} \!\big(e^{-{v^2/2\sigma^2}}-e^{-{v^2_e(r)/2\sigma^2}} \big)\>,
    \label{eq:King}
\end{equation}
generally known as the King DF, valid for a single-mass model that describe a GC. The King model provides a good match between the surface density profile of the theoretical model and the surface brightness of the observed GCs in our Galaxy \cite{King66}. 
However, the single-mass approximation does not allow a complete understanding of these systems, where the presence of a distribution of masses plays an important role, affecting the dynamical evolution of GCs and their structure, stability and observational properties. 

\section{\label{sec:StellarDynamicsMM}Stellar dynamics in a multi-mass collisional system}

The presence of a mass distribution in GCs has led some authors to study the structure of GCs using the King multi-mass models (e.g., the discrete multi-mass model of DaCosta and Freeman \cite{DF} with a measured mass function, the continuous multi-mass model by Merafina \cite{Merafina2019} using, however, an initial mass distribution \cite{kroupa2001}). These models consider stars with a given mass $m$ following a King-like DF and obtain the final distribution by integrating or summing over the masses. However, in order to describe a multi-mass collisional system, recalling that the King DF is an approximate solution of the Fokker-Planck equation taking into account stars with a unique mass $m$, we should consider stellar encounters between a test star and the mass spectrum of the field stars. 
The first indication in this direction was suggested by Michie \cite{Michie1963}. He considered the total distribution function  $f=\sum_i f_i(\bm{r},\bm{v};t)$, where $f_i$ is the DF of stars with mass $m_i$. Each $f_i$ must separately satisfy the Boltzmann equation (\ref{eq:CollBoltz}).
Michie defined the term on the right hand side of his Eq.~(1.1) as the temporal variation of the distribution function $f_i$ due to encounters between a star of mass $m_i$ with all the others. It should be emphasized that Michie did not give an explanation for the validity of the Boltzmann equation for each $f_i$, but he assumes it, including all possible differences due to the presence of a mass spectrum in the collisional term $\Gamma(f_i)$, which instead should be evaluated. 

In the present paper we present a different approach. We consider a function $f_m=f(\bm{r},\bm{v},m)$ describing the number of stars with position between $\bm{r}$ and $\bm{r}+d^3\bm{r}$, velocity between $\bm{v}$ and $\bm{v}+d^3\bm{v}$ and mass between $m$ and $m+dm$. The total DF (i.e., the distribution function of the system), defined as the number of stars in an infinitesimal volume $d^3\bm{r}d^3\bm{v}$ of the phase-space, can be obtained integrating all over the masses $f_{tot}(\bm{r},\bm{v})=\int_{\Delta m}f(\bm{r},\bm{v},m)dm\>$, where $\Delta m=[m_{min},m_{max}]$ defines an arbitrary interval of masses. 
For a collisional system, the total DF must satisfy the Boltzmann equation (\ref{eq:CollBoltz}). Then, it can be easily proved that this leads to the Boltzmann equation for $f_m$, namely
\begin{eqnarray}
    \frac{df_m}{dt}&&=\der{f_m}{t}+\bm{v}\cdot \bm{\nabla }f_m-\bm{\nabla} \varphi \cdot \der{f_m}{\bm{v}}
    =\Gamma(f_m) \>,
    \label{eq:Boltzmann_fm}
\end{eqnarray}
where we  have introduced $\Gamma(f_m)$ as the collisional term due to the totality of stellar encounters between a test star with mass $m$ and the field stars having a mass spectrum defined in the interval $\Delta m_a$.
Like in the standard single-mass theory, the collisional term is developed by assuming the local approximation and low energy exchanges (Fokker-Planck approximation). It is calculated by considering the probability that a star of mass $m$ in a given position and velocity during a close encounter suffers an change in velocity at a fixed position. Here, the assumptions are the same and the presence of a mass distribution formally does not affect the calculation of the collisional term. Therefore, the description by Chandrasekhar \cite{Chandra1943a} can still be followed, leading to an expression for $\Gamma(f_m)$ similar to Eq.~(\ref{eq:Gammaf}).
In this way, one should find the multi-mass diffusion coefficients $\langle\Delta v_i\rangle$ and $\langle\Delta v_i \Delta v_j\rangle$, generalizing ones of the standard case, taking into account the presence of a continuous spectrum in the mass of the field stars.

\section{\label{sec:DiffCoeffMM}Multi-mass Diffusion Coefficients}

The calculation of the diffusion coefficients follows the method usually presented in the literature (e.g., Sec 7.4.4 of Ref.~\refcite{BinneyTremaine}), by considering the dynamical friction as the net result of binary encounters in a homogeneous area corresponding to the central region of the cluster. This leads to the expressions for the diffusion coefficients in a binary collision between the test star with mass $m$ and the field star with mass $m_a$ (See Eqs.~(7.88) and (7.89) of Ref.~\refcite{BinneyTremaine}). This evaluation requires the knowledge of the DF of the field stars $f_a$. We can assumed a Maxwellian distribution like
\begin{equation}
    f_a(v_a,m_a)=\frac{\mathcal{A}(m_a)}{(2\pi \sigma_a^2)^{3/2}}e^{-{v_a^2/(2\sigma_a^2)}}\>,
    \label{eq:famultimass}
\end{equation}
where  $\mathcal{A}(m_a)$ is a normalization factor.
The theoretical evaluation of $\mathcal{A}(m_a)$ is still an open problem, 
however, it is clear that it is related to the central numerical density of the field stars, defined as $n_0(m_a)dm_a$, from
\begin{eqnarray}
    n_0(m_a) = \int_0^\infty f_a(v_a,m_a) d^3v_a= \mathcal{A}(m_a)\>.
\end{eqnarray}
In this way, the total central numerical density for the field stars is immediately given by
\begin{eqnarray}
    n_0 = \int_{\Delta m_a} n_0(m_a)dm_a=\int_{\Delta m_a}\mathcal{A}(m_a)dm_a\>.
\end{eqnarray}
The factor $\mathcal{A}(m_a)$ is a kind of weight function for the DF of stars with mass $m_a$.
It also contains information about the mass function of the system (see the discussion at end of Sec~ref{sec:solutionMMFokkerPlanck}). 

In Eq.~(\ref{eq:famultimass}), the one-dimensional velocity dispersion $\sigma_a$ is linked to the mass $m_a$ and the thermodynamic temperature $\theta$, since $\sigma_a^2=k\theta/m_a$; the same relation can be written for $\sigma$ substituting $m_a$ by $m$. 
Moreover, we note that in the standard case\cite{BinneyTremaine} $\sigma$ is used both in the distribution function of field stars with mass $m_a$ (instead of a more appropriate $\sigma_a$) and also as a scale for the velocity $v$ of the test star with mass $m$. 
Here, we also distinguish the thermodynamic temperature $\theta$ from the kinetic one $T$ associated with the mean square velocity of stars due to the presence of an escape velocity in the DF which limits the available phase-space \cite{Merafina2019}. 
The slight change in the DF of the field stars leads to substituting $n$ by $\mathcal{A}(m_a)$, $\rho$ by $m_a\mathcal{A}(m_a)$ and $x$ by $x_a=v/\sqrt{2}\sigma_a$ in Eq.~(7.91) and (7.92) of Ref.~\refcite{BinneyTremaine} respectively, which shows the dependence of the diffusion coefficients for binary encounters on the mass $m_a$ of the field stars.
Finally, the expressions for the multi-mass diffusion coefficients are obtained by integrating the previous ones over the range of masses $\Delta m_a$. Explicitly we have
\begin{subequations}
\begin{eqnarray}
     \langle \Delta v_i \rangle &=& \frac{v_i}{v} \langle \Delta v_{\|}\rangle\>,\\
         \langle \Delta v_i \Delta v_j \rangle&=& \frac{\delta_{ij}}{2}\langle \Delta v^2_{\bot}\rangle +\frac{v_iv_j}{v^2} \bigg(\langle \Delta v^2_{\|}\rangle
         -\frac{1}{2}\langle \Delta v^2_{\bot}\rangle\bigg)\>,\qquad\quad
\end{eqnarray}
\end{subequations}
with 
\begin{subequations}
\begin{eqnarray}
         \langle \Delta v_{\|}\rangle&=& - \int_{\Delta m_a} \frac{4\pi G^2  m_a(m+m_a)\mathcal{A}(m_a)\ln{\Lambda}}{\sigma_a^2}G(x_a)\>dm_a,\quad\qquad\\
         \langle \Delta v^2_{\|} \rangle&= & \int_{\Delta m_a} \frac{4\sqrt{2}\pi G^2 m^2_a \mathcal{A}(m_a)\ln{\Lambda}}{\sigma_a}\frac{G(x_a)}{x_a}\> dm_a\>, \\
\langle \Delta v^2_{\bot} \rangle&=& \int_{\Delta m_a}\frac{4\sqrt{2}\pi G^2 m^2_a \mathcal{A}(m_a)\ln{\Lambda}}{\sigma_a}\bigg[\frac{\erf{(x_a)}-G(x_a)}{x_a}\bigg] dm_a\>,
\end{eqnarray}
\end{subequations}
where the Coulomb logarithm $\ln{\Lambda}$ (that depends on $m_a$) and $G(x)$ are given by\cite{BinneyTremaine}
\begin{eqnarray}
 \ln{\Lambda} &=& \ln{\frac{b_{max}V_0^2}{G(m+m_a)}},\\
 G(x)&=&\frac{2}{\sqrt{\pi}x^2}\int_0^{x} y^2\eu^{-y^2}dy\>.
\end{eqnarray}

\section{\label{sec:MMFokkerPlanck}The multi-mass Fokker-Planck equation}

Since the multi-mass diffusion coefficients are obtained integrating the standard coefficients for binary encounters over the entire mass range $\Delta m_a$, the expression for $\Gamma(f_m)$ can be written in an integral form as 
\begin{equation}
    \Gamma(f_m)=\int_{\Delta m_a} \Gamma(f_m, m_a) dm_a\>.
\end{equation}
Then we obtain the expression for the binary collisional term that depends on the dimensionless variable $x_a$. We have
\begin{eqnarray}
      \Gamma(f_m,m_a) =  \frac{1}{t_R(m_a,m)}\frac{1}{x_a^2}\der{}{x_a}\bigg\{2x_aG(x_a) 
      \bigg[2x_a\frac{m}{m_a}f_m +\der{f_m}{x_a}\bigg]\bigg\}\>,
      \label{eq:Gammafm|ma_final}
\end{eqnarray}
with
\begin{eqnarray}
t_R(m_a,m) = \frac{\sqrt{2}\sigma_a^3}{\mathcal{A}(m_a)\pi G^2m_a^2\ln{\Lambda}},
\end{eqnarray}
where Eq.~(\ref{eq:Gammafm|ma_final}) is the Spitzer-H\"arm form of the Fokker-Planck equation, formally the same as the standard treatment, despite the presence of the subscript in the dimensionless velocity and the normalization term $\mathcal{A}(m_a)$. The parameter $t_R(m_a,m)$ is the relaxation time due to gravitational encounters between the two masses $m$ and $m_a$.

If we now introduce the dimensionless velocity $x={v/\sqrt{2}\sigma}$, using the definition of $x_a$ it is possible to write
\begin{eqnarray}
    \Gamma(f_m,m_a) =  \frac{1}{t_R(m_a,m)}\bigg(\frac{\sigma_a}{\sigma}\bigg)^{3}\frac{1}{x^2}\der{}{x}\bigg\{2xG(x_a)\bigg[2x\bigg(\frac{\sigma}{\sigma_a}\bigg)^2\frac{m}{m_a}f_m +\der{f_m}{x}\bigg]\bigg\}.
\end{eqnarray}
By considering the expressions for $\sigma$ and $\sigma_a$ we obtain
\begin{eqnarray}
     \Gamma(f_m,m_a)=  \frac{1}{t_R(m_a,m)}\bigg(\frac{m}{m_a}\bigg)^{\frac{3}{2}}\frac{1}{x^2}\der{}{x}\bigg\{2xG(x_a)\bigg[2xf_m +\der{f_m}{x}\bigg]\bigg\}\>.
\end{eqnarray}
When integrating over the mass range of the field stars, the dependence on $m_a$ can be grouped in the ratio
\begin{eqnarray}
    \frac{\mathcal{G}(x,m)}{t_R(m)}&=& \int_{\Delta m_a} \frac{1}{t_R(m_a,m)}\bigg(\frac{m}{m_a}\bigg)^{\frac{3}{2}}G(x_a)dm_a,
    \label{eq:Gcalligrafico}
\end{eqnarray}
where we introduced the function $\mathcal{G}(x,m)$ and the relaxation time $t_R(m)$ valid for collisions of the mass $m$ with all other masses. 
The collisional term in the Fokker-Planck equation (\ref{eq:Boltzmann_fm}) for the DF $f_m$ can now be rewritten as 
\begin{equation}
    \Gamma(f_m)=\frac{1}{t_R(m)}\frac{1}{x^2}\der{}{x}\bigg\{2x\mathcal{G}(x,m)\bigg(2xf_m +\der{f_m}{x}\bigg)\bigg\}\>,
    \label{eq:FP-SpitzerHarm_mm}
\end{equation}
which is the final form of the Fokker-Planck equation for a multi-mass collisional system. 
It should be recalled that this equation is valid only in the center of the cluster where $r\approx0$.

\section{\label{sec:solutionMMFokkerPlanck}The solution of the multi-mass Fokker-Planck equation}

We have found that reproducing the King approach \cite{King65} leads to solving Eq.~(\ref{eq:FP-SpitzerHarm_mm}). 
Let us assume that $f_m(x;t)=\exp{\big[-\lambda\> {t/t_R(m)}\big]}g(x,m)$, where $\lambda$ is the evaporation rate of the stars with mass $m$. In analogy with the standard case, we can write $g(x,m)=A(m)\Bar{g}(x,m)$, where $A(m)$ is a normalization factor. Thus the multi-mass Fokker-Planck equation becomes
\begin{eqnarray}
    &&\frac{d}{dx}\bigg\{2x\mathcal{G}(x,m)\bigg[2x\Bar{g}(x)+\frac{d\Bar{g}(x)}{dx}\bigg]\bigg\}+\lambda x^2\Bar{g}(x)=0 \>,\nonumber\\
    &&\lambda= -\frac{t_R(m)}{f_m}\frac{df_m}{dt}\>,
    \label{eq:FPmultimass}
\end{eqnarray}
where the explicit dependence on mass for $\Bar{g}(x,m)$ has been omitted. 
Similarly, using a series expansion in $\lambda$, such that $\Bar{g}(x)=\Bar{g}_0(x)+\lambda \Bar{g}_1(x)+\lambda^2\Bar{g}_2(x)+...$ with the boundary conditions $\Bar{g}(0)=1$, $\Bar{g}'(0)=0$ and $\Bar{g}(x_e)=0$, and by considering the terms with the same power of $\lambda$, the following equations must be solved
\begin{eqnarray}
        \frac{d\Bar{g}_0(x)}{dx}+2x\Bar{g}_0(x)&=& 0\>,\nonumber\\
        &&\\
        \frac{d\Bar{g}_{i+1}(x)}{dx}+2x\Bar{g}_{i+1}(x)&=& -\frac{1}{2x\mathcal{G}(x,m)}\int_0^{x}\Bar{g}_{i}(y)y^2dy\>,\nonumber
\end{eqnarray}
where $i=0,1,2...$ etc.  
The first equation, due to the boundary condition $\Bar{g}_0(0)=1$, leads to the solution $\Bar{g}_0(x)=\eu^{-x^2}$. This reproduces the Maxwell-Boltzmann dimensionless solution because $x^2={v^2/2\sigma^2}$. It is also interesting that distinguishing between $\sigma$ and $\sigma_a$ led to the vanishing of the presence of both masses $m$ and $m_a$ as in the King single mass case.  

Requiring that at zero order the time-independent part $g(x,m)$ of the DF  reproduce the Maxwell-Boltzmann distribution in Eq.~(\ref{eq:famultimass}) for the mass $m$ and also considering the dimensional factors, we obtain the relation
\begin{eqnarray}
    A(m) = \mathcal{A}(m)/(2\pi\sigma^2)^{3/2}, 
    \label{eq:A(m)multimass}
\end{eqnarray}
which therefore is similar to Eq.~(\ref{eq:A_monomass}).  
The second equation can be simplified to $\Bar{g}_{i+1}'(x)+2x\Bar{g}_{i+1}(x)=Q_i(x)$, where the boundary condition $\Bar{g}_i(0) = 0$ leads to $\Bar{g}_{i+1}(x)=e^{-x^2}\int_0^x Q_i(t)e^{t^2}dt$, with
\begin{eqnarray}
       Q_i(t)&=& -\frac{1}{2t\mathcal{G}(t,m)}\int_0^{t}\Bar{g}_i(y)y^2dy.
\end{eqnarray}

At first order $\Bar{g}(x)=\Bar{g}_0(x)+\lambda \Bar{g}_1(x)$, and it is possible to calculate the previous ratio for $i=0$
using the Michie approximation (also used by King \cite{King65}). We have
\begin{eqnarray}
        \mathcal{R} \>= \frac{\int_0^{t \sqrt{{m_a/m}}}y^2\eu^{-y^2}dy}{\int_0^ty^2 \eu^{-y^2}dy}\approx \frac{\int_0^{\infty}y^2\eu^{-y^2}dy}{\int_0^\infty y^2 \eu^{-y^2}dy}=1\>.
\end{eqnarray}
This approximation considers the function $Q_i(t)$ to be weighted by an exponential in the integration over $t$. This means that large values of $t$ are more important, while low values can be neglected. 
Consequently $\Bar{g}_1(x)$ becomes
\begin{eqnarray}
    \Bar{g}_1(x)&\approx&
   -\frac{\sqrt{\pi}}{8}({1-e^{-x^2}})\bigg[{\int_{\Delta m_a}\bigg(\frac{m}{m_a}\bigg)^{\frac{5}{2}}\frac{t_R(m)}{t_R(m_a,m)}dm_a}\bigg]^{-1}\>,
\end{eqnarray}
while, from the boundary condition $\Bar{g}(x_e)=\Bar{g}_0(x_e)+\lambda \Bar{g}_1(x_e)=0$, we obtain the expression for the evaporation rate
\begin{equation}
    \lambda = \frac{8}{\sqrt{\pi}}\bigg[\int_{\Delta m_a}\bigg(\frac{m}{m_a}\bigg)^{\frac{5}{2}}\frac{t_R(m)}{t_R(m_a,m)}dm_a\bigg] \frac{1}{e^{x_e^2}-1}\>
    \label{eq:lambdamm}
\end{equation}
and finally the expression for $\Bar{g}(x,m)$
\begin{equation}
    \Bar{g}(x,m)=\frac{e^{-x^2}-e^{-x^2_e}}{1-e^{-x^2_e}}\>,
    \label{eq:g(x)mm}
\end{equation}
where the dependence on the mass is in $x$ and $x_e$.
The resulting DF shows that considering the totality of stellar encounters with a spectrum of masses leads to the same expression for the time-independent part of the dimensionless DF for the single mass case. 

What has really changed is the expression for the evaporation rate $\lambda$, which exhibits a nontrivial dependence on its variables, in particular the relaxation time $t_R(m)$, for which an analytical expression is under development.
However, one can obtain $\lambda/t_R(m)$, which appears in the distribution function $f_m(x;t)$, by calculating the integral over $m_a$ and then make explicit the dependence in $t_R(m_a,m)$, as $\mathcal{A}(m_a)$. 
In contrast with the single mass model, the normalization term $A(m)$ depends on the mass, so some multi-mass models in the literature can be linked together by using a relation between the constants ${A}(m)$ which many authors define differently, maintaining the same physical meaning, namely a kind of weight function for each mass \cite{Sollima2017,Peuten2017,Ebrahimi2020,DF,GielesZocchi2015,Merafina2018}. 
Gathering the mass-dependent factors inside $k(m)$, such that
\begin{eqnarray}
    k(m) = \frac{A(m)}{1-e^{-x^2_e}},
    \label{eq:k(m)}
\end{eqnarray}
and integrating over the masses, then we can apply Jeans theorem to the resulting solution for $g(x)$, leading to an expression similar to Eq.~(\ref{eq:King}) as a function of the kinetic energy $\varepsilon$
\begin{equation}
   g(r,\varepsilon)=\int_{\Delta m}k(m)e^{-m[\varphi(r)-\varphi_0]/{k\theta}}\Big(e^{-{\varepsilon}/{k\theta}}-e^{-{\varepsilon_c(r)}/{k\theta}} \Big)dm\>,
   \label{eq:g(r,e,m)mm}
\end{equation}
where the kinetic energy for a star with mass $m$ is $\varepsilon=mv^2/2$, the cutoff kinetic energy is $\varepsilon_c(r)=mv_e^2(r)/2$ and as usual $\sigma^2=k\theta/m$. 

This DF can be used to study the equilibrium configurations of the system.
It also contains information about the mass distribution of the system through $\xi(m)$.
Proceeding differently, a single equilibrium configuration described by $g(r,v,m)$ will give the mass function $\xi(m)$ by integrating over positions and velocities
\begin{equation}
\xi(m)=\int_0^R \int_0^{v_e} g(r,v,m)d^3rd^3v\> ,   
\end{equation}
where $R$ is the radial edge of the system. In this way, the number of stars with mass between $m$ and $m+dm$ is $dN(m)=\xi(m)dm$, which directly shows that the total number of stars for each configuration is obtained by integrating the DF over masses, velocities and positions. Here, the normalization factor $k(m)$ is associated with $\xi(m)$ and plays an important role, if related to observations of the mass function in order to constrain the form of $\mathcal{A}(m)$. 

\section{\label{sec:}The relation between the Da Costa and Freeman factors $a_i$, the masses and the observed mass function slope}

As an example, we now discuss the relation between the factors $a_i$ in the Da Costa and Freeman discrete model, the masses and their relation with some observed mass function slopes\cite{DF}.
We recall  that their model assumes the DF of a generic mass $m_i$ is a King-like DF in terms of the energy, such that
\begin{eqnarray}
 g_i(r,v) = \alpha_i\Big[e^{-E_i/m_i \sigma_i^2}-e^{C/\sigma_i^2}\Big],
\label{eq:g_iDF}
\end{eqnarray}
valid for $E_i\leq -m_i C$, where $C>0$ is a constant related to the normalization of the gravitational potential at the edge $C=-\varphi_R$, where $R=r_t$ is the tidal radius of the system, $\sigma_i$ is related to the velocity dispersion of the stars with mass $m_i$, while $\alpha_i$ are normalization factors also related to the mass $m_i$. The energy for each mass class is given by $E_i =m_i v^2/2 + m_i\varphi(r)$. In order to obtain the density profile and solve the Poisson equation, the authors introduced the numerical factors $a_i$ defined by
\begin{eqnarray}
 a_i = \hat{\alpha}_i \exp[-\Hat{C}(1-\Hat{m}_i)],
 \label{eq:a_i}
\end{eqnarray}
where $\hat{\alpha}_i = \alpha_i/\alpha_1$, $\hat{C}=C/\sigma_1^2$ and $\hat{m}_i=m_i/m_1$ ($i=1$ refers to the higher mass class). 

The $a_i$ factors were constrained by Da Costa and Freeman using the observational properties of the M3 cluster and presented in their paper.
Our purpose is to understand how these factors depend on mass. It is clear that their values can be constrained by the observed mass function, since integrating $g_i(r,v)$ over velocities and positions leads to the number of particles in the mass class $m_i$. Since the single M3 cluster cannot alone  provide sufficient information on the functional mass dependence of the $a_i$ factors, we use the global mass function properties of Ebrahimi et.~al 2020 \cite{Ebrahimi2020}, in particular the estimation of the mass function slopes in the mass range $[0.2, 0.8]M_{\odot}$ to perform, for each of 32 GCs, a fit between the $a_i$ values and the dimensionless masses $\hat{m}_i$.

The concentrations $c$ from the Harris catalog (2010 edition) \cite{Harris1996} are used to identify the equilibrium configuration for each cluster and evaluate the parameter $W_{0,1}=m_1(\varphi_R-\varphi_0)/k\theta$, the central dimensionless gravitational potential well in the King formalism, related to the mass $m_1$ (see Ref.~\refcite{Merafina2017}, \refcite{Merafina2019}).
Here $\theta$ is the thermodynamic temperature defined in Sec.~\ref{sec:DiffCoeffMM}. We assumed that the concentrations of Harris, estimated with King single-mass models, are very close to the multi-mass concentrations $c = log(r_t/r_{k,1})$, where $r_{k,1}$ is the King radius referred to the mass $m_1$, since the surface density profiles for single and multi-mass models with the same concentrations are equal or very similar \cite{Merafina2019}. 

\begin{figure}[htbp]
    \centering
    \includegraphics[width=12cm]{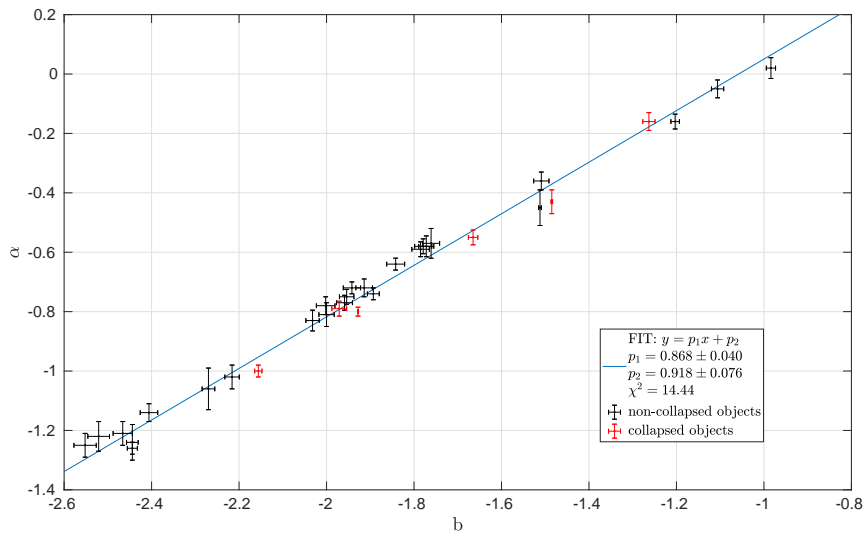}
    \caption{Correlation between the slope $\alpha$ of the observed mass function \cite{Ebrahimi2020} and the parameter $b$ of our fit between $a_i$ and $\hat{m}_i$.}
    \label{fig:alpha_b_fit}
\end{figure}

The fitting function is a power law given by $a_i = k\hat{m}_i^b + d$, in good agreement for all the clusters where the weighting factors are related to the mass function.
Table~\ref{tab:fit_ai_mi_powerlaw} lists the mass function slopes $\alpha$, the concentrations $c$, the estimated $W_{0,1}$ and the fit parameters, together with their errors. Note that, for $i=1$, it is immediately clear from the fitting function that $k+d=1$.
\begin{table}[htbp]
    \centering
     \tbl{Structural and fit parameters for 32 GCs: mass function slopes $\alpha$ by Ebrahimi et al. \cite{Ebrahimi2020}, concentrations $c$ by Harris \cite{Harris1996} (the "c" symbol denote a collapsed cluster), estimated $W_{0,1}$ and fit parameters for $a_i=k\hat{m}_i^{b}+d$, obtained from the analysis made with the Curve Fitting Toolbox in Matlab R2021a.}
{    \begin{tabular}{cccccccc}\hline
NGC	        &$\alpha$	        & $c$&	    &$W_{0,1}$  &	$k$	            &	$b$           	&	$d$	\\\hline
104	    &-0.45$\pm$ 0.12 &  2.07 	&&	11.3	&3.41	$\pm$0.03 	&-1.512	$\pm$0.006	&-2.43	$\pm$0.04\\
288	    &-0.75$\pm$	0.05&	0.99	&&	6.2		&2.25	$\pm$0.11	&-1.954	$\pm$0.033	&-1.34	$\pm$0.18\\
362	    &-0.79$\pm$	0.05&	1.76 &c&	10.3	&3.55	$\pm$0.18	&-1.971	$\pm$0.034	&-2.72	$\pm$0.30\\
1261	&-0.72$\pm$	0.04&	1.16 &	&	7.4		&2.73	$\pm$0.16	&-1.942	$\pm$0.040	&-1.87	$\pm$0.27\\
1851	&-0.74$\pm$	0.04&	1.86&  &    10.7    &3.50   $\pm$0.14   &-1.893 $\pm$0.027  &-2.63 $\pm$0.45 \\
2298    &-0.05$\pm$ 0.06&   1.38 & &	8.15	&3.93   $\pm$0.19	&-1.106	$\pm$0.028	&-2.98	$\pm$0.23\\
3201	&-1.22$\pm$	0.10&	1.29&	&	8.6		&3.12	$\pm$0.22	&-2.521	$\pm$0.050	&-2.51	$\pm$0.53\\
4590	&-1.25$\pm$	0.08&	1.41&	&	9.2		&3.37	$\pm$0.25	&-2.552	$\pm$0.051	&-2.83	$\pm$0.60\\
5024	&-1.21$\pm$	0.08&	1.72&	&	10.6	&3.70	$\pm$0.23	&-2.466	$\pm$0.044	&-3.11	$\pm$0.54\\
5053	&-1.26$\pm$	0.04&	0.74&	&	4.4		&1.60	$\pm$0.05	&-2.444	$\pm$0.023	&-0.68	$\pm$0.12\\
5272	&-1.02$\pm$	0.08&	1.89&	&	11.1	&3.51	$\pm$0.17	&-2.216	$\pm$0.033	&-2.74	$\pm$0.32\\
5286	&-0.64$\pm$	0.04&	1.41&	&	8.7		&3.30	$\pm$0.20   &-1.842	$\pm$0.041	&-2.46	$\pm$0.32\\
5466	&-1.14$\pm$	0.06&	1.04&	&	6.9		&2.39	$\pm$0.14	&-2.406	$\pm$0.040	&-1.60	$\pm$0.31\\
5897	&-1.06$\pm$	0.14&	0.86&	&	5.4		&1.89	$\pm$0.08	&-2.270	$\pm$0.029	&-0.99  $\pm$0.16\\
5904	&-0.81$\pm$	0.08&	1.73&	&	10.2	&3.55	$\pm$0.19	&-2.000	$\pm$0.035	&-2.74	$\pm$0.32\\
5986	&-0.58$\pm$	0.05&	1.23&	&	7.7		&2.94	$\pm$0.17	&-1.779	$\pm$0.039	&-2.06	$\pm$0.27\\
6093	&-0.16$\pm$	0.05&	1.68&	&	9.5		&4.09	$\pm$0.14	&-1.203	$\pm$0.020	&-3.14	$\pm$0.17\\
6101	&-1.24$\pm$	0.12&	0.80&	&	4.9		&1.72	$\pm$0.07	&-2.444	$\pm$0.027	&-0.82  $\pm$0.15\\
6144	& 0.02$\pm$	0.07&	1.55&	&	8.85	&4.49	$\pm$0.18	&-0.984	$\pm$0.021	&-3.53	$\pm$0.20\\
6218	&-0.36$\pm$ 0.06&	1.34&	&	10		&3.31	$\pm$0.18	&-1.509	$\pm$0.035	&-2.41	$\pm$0.25\\
6254	&-0.57$\pm$	0.10&	1.38&	&	8.5		&3.27	$\pm$0.19	&-1.761	$\pm$0.039	&-2.41	$\pm$0.30\\
6341	&-0.77$\pm$	0.05&	1.68&	&	10		&3.54	$\pm$0.19	&-1.959	$\pm$0.037	&-2.73	$\pm$0.32\\
6362	&-0.58$\pm$	0.07&	1.09&	&	6.8		&2.56	$\pm$0.13	&-1.772	$\pm$0.035	&-1.66	$\pm$0.21\\
6541	&-0.55$\pm$	0.05&	1.86 &c&	10.5	&3.56	$\pm$0.12	&-1.665	$\pm$0.022	&-2.65	$\pm$0.17\\
6584	&-0.78$\pm$	0.06&	1.47&	&	9.1		&3.37	$\pm$0.21	&-2.002	$\pm$0.043	&-2.58	$\pm$0.37\\
6723	&-0.16$\pm$	0.06&	1.11& c&	6.8		&3.00	$\pm$0.14   &-1.263	$\pm$0.028	&-2.05	$\pm$0.17\\
6752	&-0.43$\pm$	0.08&	2.50 &c&	13.4	&2.96	$\pm$0.02	&-1.485	$\pm$0.004	&-1.95	$\pm$0.02\\
6779	&-0.59$\pm$	0.05&	1.38&	&	8.5		&3.26	$\pm$0.20	&-1.785	$\pm$0.040	&-2.40	$\pm$0.30\\
6809	&-0.83$\pm$	0.07&	0.93&	&	5.9		&2.08	$\pm$0.09	&-2.032	$\pm$0.031	&-1.17	$\pm$0.16\\
7078	&-1.00$\pm$	0.04&	2.29& c&	12.8	&3.03	$\pm$0.07	&-2.156	$\pm$0.017  &-2.12	$\pm$0.14\\
7089	&-0.72$\pm$ 0.06&  1.59 &	&	9.6		&3.50	$\pm$0.20	&-1.914	$\pm$0.038	&-2.69	$\pm$0.33\\
7099	&-0.8$\pm$	0.03&	2.50 &c&	13.7	&2.84	$\pm$0.04	&-1.928	$\pm$0.001	&-1.87	$\pm$0.07\\\hline
\end{tabular}\label{tab:fit_ai_mi_powerlaw}}
\end{table}
We also found there is a linear correlation between the mass function slope $\alpha$ and the exponent $b$. The fit results with the errors at 95\% confidence level and the $\chi^2$-test value are reported in Figure~\ref{fig:alpha_b_fit}, where we also distinguish the normal GCs from those which are collapsed.
The linear fitting procedure $\alpha = p_1 b+p_2$ gives for the parameters $p_1 = 0.868\pm0.040$ and $p_2=0.918 \pm 0.076$, with $\chi^2=14.44$. 

\begin{figure}[htbp]
    \centering
    \includegraphics[width=12cm]{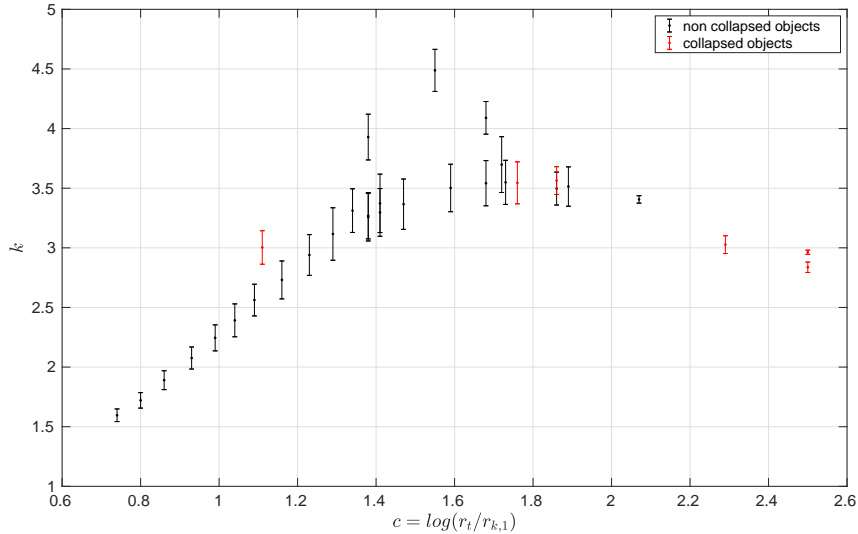}
    \caption{Correlation between the parameter $k$ and the concentration $c$ \cite{Harris1996} for the 32 analyzed GCs.}
    \label{fig:k_conc}
\end{figure}

There is also a correlation between the parameter $k$ and the concentration $c$, as shown in Figure~\ref{fig:k_conc}, even if such a correlation becomes less clear at large values of $c$. 
The figure suggests an increasing linear behavior for $c<1.4$.
Note that large values of $c$ correspond, typically, to collapsed clusters.

A deeper look into the normalization of the DF in Da Costa and Freeman model shows that the factors $\alpha_i$ and the constants $a_i$ can be related to the factor $k(m)$ in Eq.~(\ref{eq:g(r,e,m)mm}).
Indeed writing the DF $g_i$ of Eq.~(\ref{eq:g_iDF}) in terms of the kinetic energy, the cut-off kinetic energy and the gravitational potential, it is easy to show the validity of the following relation between $a_i$ and $k_i$ (i.e., the discrete extrapolation of our continuous factor $k(m)$ in the DF)
\begin{equation}
    a_i=\hat{k}_ie^{W_{0,1}-W_{0,i}},
\end{equation}
where $\hat{k}_i=k_i/k_1$ and $W_{0,i}=\hat{m}_iW_{0,1}=m_i(\varphi_R-\varphi_0)/k\theta$ and a fixed range for each mass class was considered (i.e. each $k_i$ is valid in a mass class with range $\Delta m_i$ that was assumed equal for each class). Therefore, the possibility of evaluating the $k_i$ terms is linked to being able to find the values of the parameter $W_{0,1}$ and the scaling factor $k_1$. This is  in principle possible if the central numerical density for the heaviest mass $m_1$ is known from observations in the inner regions of GCs and an evaluation of the number of the heavy stars of class $m_1$.
The evaluation of $k(m)$ from observational data can be considered very tricky and will be addressed to a forthcoming paper.

\section{\label{sec:conclusions}Conclusions}

In this paper we have presented an improvement on the theory of stellar orbit diffusion in multi-mass collisional systems having a continuous mass distribution. 
A new expression is derived for the diffusion coefficients considered in the Fokker-Planck equation for the DF of a generic test star with mass $m$. 
Following the approach by King in 1965 \cite{King65}, a steady-state solution for the DF in a King-like form \cite{King66} is found, while the expression for the evaporation rate is slightly different.
This encourages the basic assumption of several multi-mass models for GCs found in the literature. 
In our formulation, the DF of the test stars also gives information about the mass function of the system, which changes slightly the form of the DF. 
In the analysis of the Da Costa and Freeman discrete multi-mass model, we considered 32 GCs for which the global mass function slope was estimated in the mass range $[0.2, 0.8]M_{\odot}$ by Ebrahimi et al.\cite{Ebrahimi2020}. The knowledge of the mass function, together with the estimation of their concentration through the Harris catalog (2010 edition) \cite{Harris1996}, led us to find a correlation between the weight factors $a_i$ and the dimensionless masses by the relation $a_i=k\hat{m}_i^b+d$. A linear relation between the parameter $b$ and the estimated slope $\alpha$ has been obtained as well a correlation between the parameter $k$ and the concentration $c$. Finally, we presented the relation between the weighting factors of the DaCosta and Freeman model and ours, directly arising from the solution of the derived multi-mass Fokker-Planck equation. 

Understanding the dynamical evolution of GCs through the development of multi-mass models is in continuous improvement even if some aspects remain open problems. 
In particular, the meaning of the factor $\mathcal{A}(m)$, its dependence on mass and its impact on the quantities which depend on it requires further study.
This factor appears in all the physical quantities that can be predicted by the model, such as the brightness profile, the mass function, the velocity dispersion, the evaporation rate and the mass segregation effect, a phenomenon that tends to sink heavy stars and bring the lightest ones toward the outer regions. 
A better understanding of the role of the mass function in the dynamical evolution and the thermodynamic instabilities present in GCs is required and it is still under discussion (see e.g. Refs.~\refcite{Merafina2017}, \refcite{Merafina2019}). 
Finally, the recent possibility of measuring the transverse velocity of the stars in GCs allows the analysis of the internal kinematic and test multi-mass models using dynamical quantities, beyond the structural parameters and brightness profiles. This possibility enables the verification of the possible phenomenon of equipartition in the presence of mass segregation \cite{Parker2016}, currently not yet well understood.

\nocite{*}

\bibliographystyle{ws-ijmpd}
\bibliography{sample.bib}

\end{document}